\documentclass[a4paper,10pt,twocolumn]{article}
\usepackage{newtxtext,newtxmath}
\usepackage{amsmath}
\usepackage{upgreek}
\usepackage{graphicx}
\usepackage[margin=15mm,a4paper]{geometry}
\usepackage[final]{hyperref}
\hypersetup{colorlinks=true,linkcolor=blue,citecolor=blue,filecolor=magenta,urlcolor=blue}

\title{\Large\textbf{Calibration of the surface array of the Pierre Auger Observatory}}
\author{\normalsize
X.~Bertou,$^\text{a}$
P.\,S.~Allison,$^\text{b}$\footnotemark~~
C.~Bonifazi,$^\text{c}$
P.~Bauleo,$^\text{d}$
C.M.~Grunfeld,$^\text{e}$
M.~Aglietta,$^\text{f}$
F.~Arneodo,$^\text{g}$
D.~Barnhill,$^\text{h}$
\\\normalsize
J.J.~Beatty,$^\text{b}$
N.G.~Busca,$^\text{i,j,k}$
A.~Creusot,$^\text{l}$
D.~Dornic,$^\text{l}$
A.~Etchegoyen,$^\text{m}$
A.~Filevitch,$^\text{m}$
P.L.~Ghia,$^\text{f,g}$
I.~Lhenry-Yvon,$^\text{l}$
\\\normalsize
M.C.~Medina,$^\text{m}$
E.~Moreno,$^\text{n}$
D.~Nitz,$^\text{o}$
T.~Ohnuki,$^\text{h}$
S.~Ranchon,$^\text{p}$
H.~Salazar,$^\text{n}$
T.~Suomij\"arvi,$^\text{l}$
D.~Supanitsky,$^\text{m}$
\\\normalsize
A.~Tripathi,$^\text{h}$
M.~Urban,$^\text{p}$
and L.~Villasenor,$^\text{q}$
for the Pierre Auger Collaboration\footnotemark
\\[2mm]\footnotesize
$^\text{a}$\emph{Centro At\'omico Bariloche (CNEA), S.C.~de Bariloche, Argentina}
\\[-1mm]\footnotesize
$^\text{b}$\emph{Department of Physics, Ohio State University, 191 W.~Woodruff Ave., Columbus, OH 43201, USA}
\\[-1mm]\footnotesize
$^\text{c}$\emph{CBPF/IN2P3-CNRS, Rua Xavier Sigaud, 150 Rio de Janeiro, Brazil}
\\[-1mm]\footnotesize
$^\text{d}$\emph{Department of Physics, Colorado State University, Fort Collins, CO 80523, USA}
\\[-1mm]\footnotesize
$^\text{e}$\emph{Universidad Nacional de la Plata, Facultad de Ciencias Exactas, Departamento de F{\'\i}sica and IFLP/CONICET, C.C.~67, (1900) La Plata, Argentina}
\\[-1mm]\footnotesize
$^\text{f}$\emph{Istituto di Fisica dello Spazio Interplanetario, INAF, and INFN, Torino, Italy}
\\[-1mm]\footnotesize
$^\text{g}$\emph{INFN, Laboratori Nazionali del Gran Sasso, Assergi, Italy}
\\[-1mm]\footnotesize
$^\text{h}$\emph{University of California, Los Angeles (UCLA), USA}
\\[-1mm]\footnotesize
$^\text{i}$\emph{Department of Astronomy \& Astrophysics, The University of Chicago, Chicago, IL 60637-1433, USA}
\\[-1mm]\footnotesize
$^\text{j}$\emph{Kavli Institute of Cosmological Physics, The University of Chicago, Chicago, IL 60637-1433, USA}
\\[-1mm]\footnotesize
$^\text{k}$\emph{Fermi National Accelerator Laboratory, Particle Astrophysics Center, Batavia, IL 60510-0500, USA}
\\[-1mm]\footnotesize
$^\text{l}$\emph{Institut de Physique Nucl\'eaire d'Orsay, Universit\'e Paris-Sud et IN2P3-CNRS, 91406 Orsay Cedex, France}
\\[-1mm]\footnotesize
$^\text{m}$\emph{Laboratorio Tandar, Comisi\'on Nacional de Energ{\'\i}a At\'omica and CONICET, Av.~Gral.~Paz 1499 (1650) San Mart{\'\i}n, Buenos Aires, Argentina}
\\[-1mm]\footnotesize
$^\text{n}$\emph{Benem\'erita Universidad Aut\'onoma de Puebla (BUAP), Ap.~Postal J-48, 72500 Puebla, Puebla, Mexico}
\\[-1mm]\footnotesize
$^\text{o}$\emph{Physics Department, Michigan Technological University, Houghton, MI 49931, USA}
\\[-1mm]\footnotesize
$^\text{p}$\emph{Laboratoire de l'Acc\'elerateur Lin\'eaire, IN2P3-CNRS et Universite Paris-Sud, Centre Scientifique d'Orsay, Bat 200, B.P.~34, 91898 Orsay Cedex, France}
\\[-1mm]\footnotesize
$^\text{q}$\emph{University of Michoacan, Morelia, Michoacan, Mexico}
}
\date{}

\begin{document}

\twocolumn[
\begin{@twocolumnfalse}
\maketitle
\vspace{-7mm}
~\\\hrule
\vspace{-1mm}
\begin{abstract}
The Pierre Auger Observatory is designed to study cosmic rays of the highest energies (${>}10^{19}$\,eV).
The ground array of the Observatory will consist of 1600 water-Cherenkov detectors deployed over 3000\,km$^2$.
The remoteness and large number of detectors require a robust, automatic self-calibration procedure.
It relies on the measurement of the average charge collected by a photomultiplier tube from the Cherenkov light produced by a vertical and central through-going muon, determined to $5-10$\% at the detector via a novel rate-based technique and to 3\% precision through analysis of histograms of the charge distribution.
The parameters needed for the calibration are measured every minute, allowing for an accurate determination of the signals recorded from extensive air showers produced by primary cosmic rays.
The method also enables stable and uniform triggering conditions to be achieved.

~


~


\end{abstract}
\hrule
\vspace{5mm}

Published in Nucl.~Instrum.~Meth.~A as \href{https://doi.org/10.1016/j.nima.2006.07.066}{DOI:10.1016/j.nima.2006.07.066}
\\
Report number: FERMILAB-PUB-21-033-AD-AE-E-TD
\vspace{10mm}
\end{@twocolumnfalse}
]

\let\theorigfootnote\thefootnote
\renewcommand{\thefootnote}{*}\footnotetext{Corresponding author, tel. $+$1 614 247 8190, \texttt{barawn@auger.org.ar}}
\renewcommand{\thefootnote}{$\dagger$}\footnotetext{Pierre Auger Observatory, Av.~San Mart{\'\i}n Norte 304, (5613) Malarg\"ue, Argentina.}
\let\thefootnote\theorigfootnote

\section{Introduction}

At energies above $10^{19}$\,eV, the cosmic ray flux is very low (${\sim}1$\,km$^{-2}$\,sr$^{-1}$\,yr$^{-1}$) requiring a very large area, sparse, simple, and reliable design, and the ability to identify rare shower candidates from a large background.
The Pierre Auger Observatory is a hybrid device optimized for energies above $10^{19}$\,eV consisting of an air fluorescence detector as well as a surface detector (SD) using water-Cherenkov tanks.
The water-Cherenkov detector method has been used prominently in previous cosmic ray air shower experiments (e.g.\ Ref.~[1]). 
The SD obtains a measurement of the Cherenkov light produced by shower particles passing through the detector at ground, and reconstructs the air shower by fitting the observed signal as a function of lateral distance from the shower core.
The Cherenkov light is measured in units of the signal produced by a vertical and central through-going (VCT) muon, termed a vertical-equivalent muon (VEM).

The SD consists of 1600 water tanks, with 10\,m$^2$ water surface area and 1.2\,m water height and three 9\,in Photonis XP1805PA/1 photomultiplier tubes (PMTs) looking into a Tyvek\textsuperscript{\textregistered} reflective liner through optical coupling material~[2].
The signal from the 3 PMTs is digitized by local electronics, and the data are sent to a central data acquisition system (CDAS) when requested.

The total bandwidth available from each SD to the CDAS is approximately 1200 bits per second~[3] which implies that the calibration must be done by the local electronics.
The local processor is an 80\,MHz PowerPC 403GCX lacking floating-point hardware, which forces the calibration to be as simple as possible.
Finally, the remoteness of the detectors implies that the calibration procedure must be robust, accepting the possibility of failures of individual PMTs, to allow for recovery of these stations in data analysis.

The SD electronics uses six 40\,MHz AD9203 10-bit flash analog-to-digital converters (FADCs) to digitize the signals from the 3 PMTs.
Two signals are taken from each PMT -- one directly from the anode, and the other from the last dynode, amplified and inverted by the base electronics to a total of nominally 32 times the anode.
The two signals are used to provide enough dynamic range to cover with good precision both the particle flux near the shower core (${\sim}1000$ particles per $\upmu$s) and far   from the shower core (${\sim}1$ particle per $\upmu$s).
The two signals are called simply the anode and dynode, respectively.
The signal recorded by the FADC is referred to in units of channels (ch), with a range of $0-1023$, corresponding to an input range of $0-2$\,V.
Each FADC bin corresponds to 25\,ns.

\section{The vertical-equivalent muon}

The primary signal calibration information required from the SD is the average charge measured for a VCT muon, named the vertical-equivalent muon (VEM, or $Q_\text{VEM}$ when needed for clarity).
During shower reconstruction, the signal recorded by the tanks is converted into units of VEM, and the shower characteristics, i.e.\ total energy and arrival direction, are fit using a lateral distribution function and energy conversion based upon hybrid analysis using the florescence detector.
The conversion to units of VEM is done either to provide a common reference level between tanks or to calibrate against the detector simulations for other Monte Carlo-based studies.
Therefore, the goal of the calibration is to obtain the value of 1\,VEM in electronics units (i.e.\ integrated channels).
 
In addition, to maintain a uniform trigger condition for the array, the station must also be able to set a common trigger threshold in detector-independent units.
This will allow for a tank-independent analysis of the acceptance of the array by modeling the trigger~[4].

There are several quantities which are strongly related to a VEM, but are determined with different methods.
These quantities are listed in Table~1 for easy reference.

~

{\noindent\small
Table 1
\\
\noindent
Reference for calibration terms\hfill
\\[3mm]
\begin{tabular}{lll}
\hline
\small
Symbol & Definition & Section
\\ 
\hline
VEM or $Q_\text{VEM}$ & Charge deposited in PMT by light & 2
\\ 
 & from VCT muon
\\
$Q^\text{peak}_\text{VEM}$ & Peak in a charge histogram & 2.1
\\
$Q^\text{est}_\text{VEM}$ & On-line estimated value of $Q^\text{peak}_\text{VEM}$ & 3.2
\\ 
$I_\text{VEM}$ & Pulse height of light from VCT muon & 2.2
\\
$I^\text{peak}_\text{VEM}$ & Peak in a pulse height histogram & 2.2
\\ 
$I^\text{est}_\text{VEM}$ & On-line estimated value of $I^\text{peak}_\text{VEM}$ & 3.2
\\ 
\hline
\end{tabular}}

\subsection{Charge histograms and their relation to a VEM}

Atmospheric muons passing through the detector at a rate of approximately 2500\,Hz give an excellent method for measuring 1\,VEM precisely, but the SD in its normal configuration has no way to select only VCT muons.
However, the distribution of the light from atmospheric muons also produces a peak in a charge distribution~[5].
This peak ($Q^\text{peak}_\text{VEM}$) is at approximately 1.09\,VEM for the sum of the 3 PMTs and $(1.03\pm0.02)$\,VEM for each PMT, both measured with a muon telescope providing the trigger in a reference tank~[6].\footnote{Based on more recent data from Ref.~[6] and a similar setup with a different tank.}
The difference between these two cases is due to the fact that the sum of the PMTs measures the total signal, whereas the individual PMTs primarily measure the portion of the signal deposited closest to them.
Example charge and pulse height histograms produced by a SD are shown in Fig.~1.
The peak produced by the response of atmospheric muons in the tank is clearly visible.
The relation of $Q^\text{peak}_\text{VEM}$ to $Q_\text{VEM}$ can be understood using a simple geometrical model~[7].
The shift observed is caused by the convolution of photoelectron statistics on an asymmetric peak in the track length distribution and local light collection effects.

\begin{figure*}[t]
\centering
\includegraphics[width=0.65\textwidth]{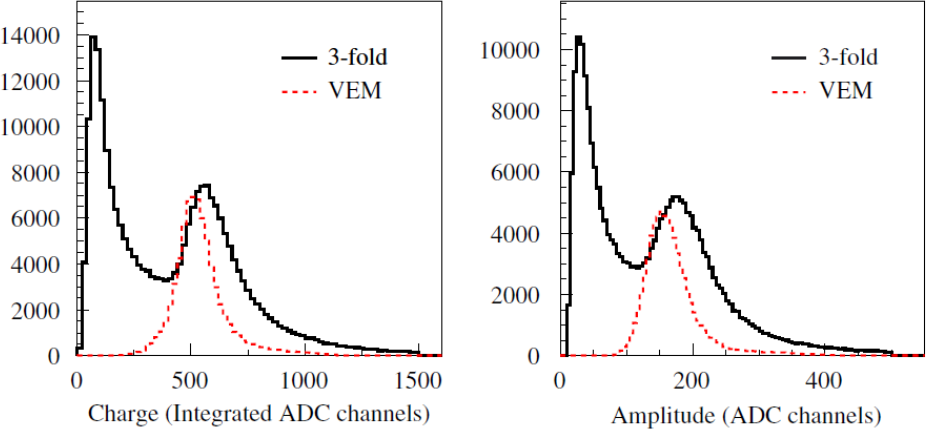}
\caption{Charge and pulse height histograms from an SD station, triggered by a 3-fold coincidence between all 3 PMTs at a trigger level of five channels above baseline, with the signal from all 3 PMTs summed.
The dashed histogram is produced by an external muon telescope providing the trigger to select only vertical and central muons.
The first peak in the black histogram is caused by the convolution of the trigger on a steeply falling distribution from low energy particles.
The second peak is due to vertical through-going atmospheric muons~[6].}
\end{figure*}

\subsection{Pulse height histograms and their relation to the trigger}

The SD uses two triggers to identify shower  candidates~[8].
The first is a simple threshold trigger, which is satisfied when the signal from all 3 PMTs exceeds a set threshold, and is designed for signals close to the shower core.
The second is a time-over-threshold trigger, which requires that the signal from 2 of the 3 PMTs exceed a much lower threshold than the first trigger for a number of time bins within a given time window, and is designed for signals far from the shower core.
These triggers are set in electronics units (channels) -- a measure of the current from the PMT -- so the station must have a reference unit for current as well.
Atmospheric muons again provide this reference, as the same mechanism (see Section~2.1) that produces a
 peak in the charge histogram also produces a peak in a pulse height histogram ($I^\text{peak}_\text{VEM}$), which is then used as the common reference unit for threshold levels.
This peak, like the charge histogram peak, is related to the peak current produced by a vertical through-going muon ($I_\text{VEM}$).

The target trigger threshold is $3.2I^\text{peak}_\text{VEM}$ for the simple threshold trigger, and $0.2I^\text{peak}_\text{VEM}$ for the time-over-threshold trigger.

The conversion from electronics units to $I^\text{peak}_\text{VEM}$ must be continually updated in order to maintain the proper trigger level.
The accuracy of this determination does not have to be high -- the trigger units are quantized in channels -- and the target trigger level of $0.2I^\text{peak}_\text{VEM}$  ($\simeq$10\,ch -- see Section~3) for the lower of the two triggers implies that the precision of the on-line calibration does not need to be much better than 10\% (1 part in 10 channels) before the quantization of the trigger dominates.

In addition, the initial end-to-end gains of the 3 PMTs -- that is, $I^\text{peak}_\text{VEM}$ -- must be roughly equivalent.
This ensures that the signals recorded from the PMTs are similar in amplitude, and sets the proper dynamic range and signal size for the electronics.

\section{VEM calibration procedure}

There are three main steps to the calibration to VEM units.
\begin{itemize}
\setlength\itemsep{-1mm}
\item[(1)] Set up the end-to-end gains of each of the 3 PMTs to have $I^\text{peak}_\text{VEM}$ at 50\,ch.
\item[(2)] Continually perform a local calibration to determine $I^\text{peak}_\text{VEM}$ in channels to adjust the electronics-level trigger.
This compensates for drifts which occur after step \#1.
\item[(3)] Determine the value of $Q^\text{peak}_\text{VEM}$ to high accuracy using charge histograms, and use the known conversion from $Q^\text{peak}_\text{VEM}$ to 1\,VEM to obtain a conversion from the integrated signal of the PMT to VEM units.
\end{itemize}

\subsection{End-to-end gain setup}

The end-to-end gains (i.e.\ $I^\text{peak}_\text{VEM}$ in electronics units) of each of the 3 PMTs are set up by matching a point in the spectrum to a measured rate from a reference tank (see Ref.~[6]).
The reference tank is calibrated by obtaining a charge histogram and adjusting the PMT high voltage until the peak ($Q^\text{peak}_\text{VEM}$) of the three histograms agree.
The singles rate spectrum of each of the PMT (i.e.\ no coincidence between the PMTs required) is then obtained as a reference.
A point on the spectrum convenient as a trigger threshold was chosen as a target calibration point for all tanks -- that is the singles rate of a PMT at 150\,ch above baseline was required to be 100\,Hz, which corresponds to a trigger point of roughly $3I^\text{peak}_\text{VEM}$.
This choice sets up each of the PMTs to have approximately 50\,ch/$I^\text{peak}_\text{VEM}$.

When the local station electronics is first turned on, each of the 3 PMTs is forced to satisfy this condition by adjusting the high voltage until the rate is 100\,Hz at a point 150\,ch above baseline.
This balances the PMTs to approximately 5\%.
The PMTs have a range of temperature coefficients so that subsequent drifts from the initial settings are inevitable.
Thus, high precision is not required.
For a sample of 661 tanks, the mean RMS spread in $I^\text{peak}_\text{VEM}$ between the 3 PMTs was 4.4\%.

The end-to-end gain measurement implies that the PMTs in the SDs will \emph{not} have equivalent gains -- indeed, even PMTs within the same tank may not have equivalent gains.
If a water tank produces more photons per vertical muon than an average tank, then the PMTs in the tank will be at a lower gain than an average tank to compensate.
Likewise, if a PMT has a worse optical coupling than the others in the same tank, resulting in fewer photons seen per vertical muon, the PMT will be run at a higher gain.
Thus, we would expect to see an inverse relationship between the gain of each PMT and the number of photoelectrons ($n_\text{pe}$) per VEM for all the PMTs in the SD.
This is shown in Fig.~2.
The inverse relationship is quite clear, showing that the initial end-to-end gain setup is operating correctly.
The choice of 50\,ch/$I^\text{peak}_\text{VEM}$ results in a mean gain of approximately $3.4{\times}10^5$ for a mean $n_\text{pe}/\text{VEM} \sim 94$\,pe.

\begin{figure*}[t]
\centering
\includegraphics[width=0.6\textwidth]{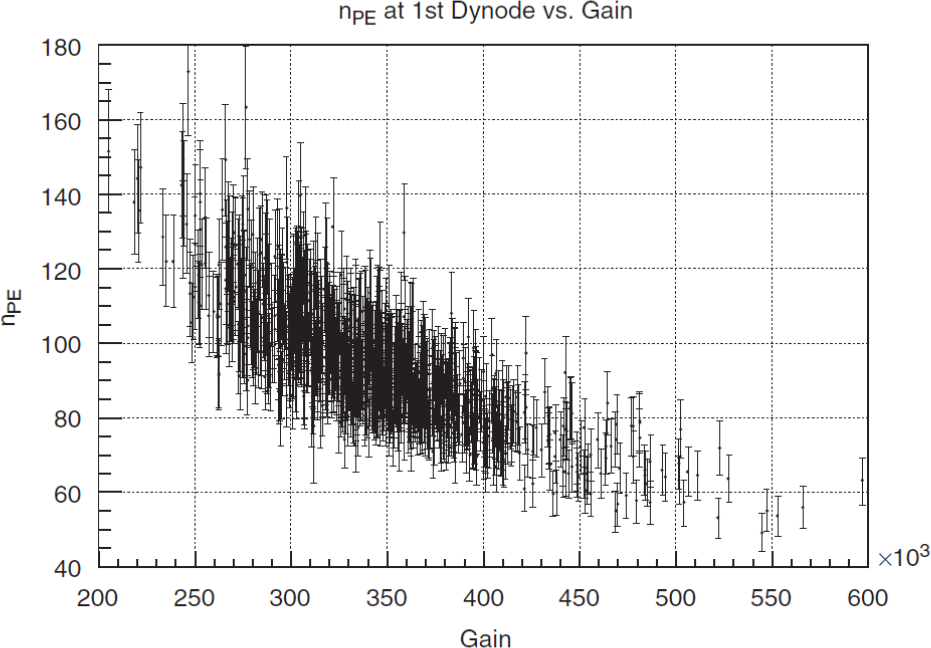}
\caption{Number of photoelectrons ($n_\text{pe}$) at the first dynode versus PMT gain.
As the end-to-end gain setup is designed to produce equivalent FADC channels for a charge deposition of 1\,VEM, an inverse relationship is expected between gain and $n_\text{pe}$.
$n_\text{pe}$ at the first dynode is calculated as described in Section~3.2.
The mean gain for the PMTs in the SD is $3.4{\times}10^5$, and the mean $n_\text{VEM}$/VEM is 94\,pe.}
\end{figure*}

\subsection{Continual on-line calibration}

Once the gains of the 3 PMTs are set up, the drifts of the value of $I^\text{peak}_\text{VEM}$ in electronics units for each detector must be compensated to ensure that the surface array triggers uniformly.
This compensation is done via adjusting the trigger levels based on a continual on-line calibration.
The PMT high voltage is not changed during normal operation, which implies that the dynamic range of the SD will be slightly non-uniform.
For normal operation, this non-uniformity is minimal ($\sim$10\%).
Detectors which have drifted significantly (${>}20$\,ch) from the nominal $I^\text{peak}_\text{VEM}$ of 50\,ch are re-initialized following the procedure in Section~3.1.
The average value of $I^\text{peak}_\text{VEM}$ for the PMTs of the SD is currently (46 $\pm$ 4)\,ch. 

The value of $I^\text{peak}_\text{VEM}$ as defined in Section~2.2 is not obtained on-line since this would increase the dead time of the detector to unacceptable levels.
Instead, the trigger levels are set with respect to an estimate of $I^\text{peak}_\text{VEM}$.
This estimate ($I^\text{est}_\text{VEM}$) is defined implicitly for a given PMT by requiring that the rate of events satisfying a ``calibration trigger'' be 70\,Hz.
An event satisfies the calibration trigger if the signal is above $2.5I^\text{est}_\text{VEM}$ for the given PMT \emph{and} above $1.75I^\text{est}_\text{VEM}$ for all three PMTs.
The value of the rate (70\,Hz) was obtained from the reference tank.

To obtain the value of $I^\text{est}_\text{VEM}$, a $\sigma$-$\delta$ convergence algorithm is used, where a test value ($I^\text{est}_\text{VEM}$) is altered by an adjustment $\delta$ if a measured value (the rate) is outside of a bound $\sigma$.
This algorithm is implemented as follows:
\begin{itemize}
\setlength\itemsep{-1mm}
\item[(1)] Start with a value of $I^\text{est}_\text{VEM}=50$\,ch.
\item[(2)] Measure, for each PMT, the rate of events satisfying the calibration trigger by counting these events for a time $t_\text{cal}$, initially set to 5\,s.
\item[(3)] If, for a given PMT, the rate is above $70 + \sigma$\,Hz, increase $I^\text{est}_\text{VEM}$ by $\delta$.
Likewise, if the rate is below $70 - \sigma$\,Hz, decrease $I^\text{est}_\text{VEM}$ by $\delta$, with $\sigma =2$\,Hz and $\delta = 1$\,ch initially.
\item[(4)] If the rate of any single PMT is more than 10\,$\sigma$ away from 70\,Hz, adjust $I^\text{est}_\text{VEM}$ by 5\,ch in the appropriate direction, set $t_\text{cal}$ to 10\,s, $\delta=1$\,ch, and repeat from step (2).
\item[(5)] Otherwise, if $t_\text{cal}<60$\,s, increase $t_\text{cal}$ by 5\,s.
If $\delta>0.1$\,ch, decrease $\delta$ by 0.1\,ch, and repeat from step (2).
\end{itemize}

As the calibration trigger is a single PMT trigger within a 3-fold coincidence, a small drift in the calibration trigger rate of one PMT should not affect the rate of another.
Step (4) allows the algorithm to switch back to a coarser tracking mode to minimize the effect that one PMT can have on the other two.
In practice, changes of less than 10\% in a period of $t_\text{cal}$ do not affect the other PMTs significantly.

A minimum value of $\delta = 0.1$\,ch was chosen to allow $I^\text{est}_\text{VEM}$ to compensate for drifts in the baseline of each channel as small as 0.1\,ch without significantly affecting the error in the estimate.
The value of $\sigma = 2$\,Hz is equal to $\sim2\times$ the Poisson fluctuation over that time period, and is reasonable given the requirement of ${<}10\%$ accuracy.

A simple test of the success of the convergence algorithm is to look at the trigger rates for the simple threshold trigger, which is just a 3-fold coincidence trigger at $3.2I^\text{peak}_\text{VEM}$.
On a reference tank, with 3 PMTs tuned to equal $I^\text{peak}_\text{VEM}$ values, this gives a rate of ${\sim}20$\,Hz.
The 3-fold coincidence rates for 21 tanks before and after the convergence algorithm is applied is shown in Fig.~3.
The rapid convergence to ${\sim}20$\,Hz shows that the method described enables the uniform trigger levels to be established rapidly.

A comparison of the converged $I^\text{est}_\text{VEM}$ value with values obtained from a pulse height histogram gives $I^\text{est}_\text{VEM}=(0.94\pm0.06)I^\text{peak}_\text{VEM}$.
The systematic offset is due to a slight inaccuracy in the required calibration trigger rate (i.e.\ $2.5I^\text{peak}_\text{VEM}$ results in 70\,Hz) and is unimportant.

The on-line calibration also estimates $Q^\text{peak}_\text{VEM}$ as well ($Q^\text{est}_\text{VEM}$) by computing the charge of pulses with a peak of exactly $I^\text{est}_\text{VEM}$, and using a $\sigma$-$\delta$ convergence algorithm on $Q^\text{est}_\text{VEM}$, determined initially from the charge of the first pulse.
A comparison of the converged $Q^\text{est}_\text{VEM}$ and $Q^\text{peak}_\text{VEM}$ determined from a peak fit to the charge histograms yields $Q^\text{est}_\text{VEM}=(0.96\pm0.03)Q^\text{peak}_\text{VEM}$.
During operation, $Q^\text{est}_\text{VEM}$ is used to monitor the status of the detector continuously and to provide a cross-check on the $Q_\text{VEM}$ histogram measurement.
Since $Q_\text{VEM}$ is just the number of photoelectrons per muon ($n_\text{pe})$ times the PMT gain, the dynode/anode ratio, and the electronic gain, $Q^\text{est}_\text{VEM}$ can be used to calculate $n_\text{pe}$ for all detectors as well.

The history over the last $7t_\text{cal}$ (60\,s) periods of the adjustments to $I^\text{est}_\text{VEM}$ is included with each event, along with $I^\text{est}_\text{VEM}$, $Q^\text{est}_\text{VEM}$, and the last 70\,Hz rates for each of the 3 PMTs.

\begin{figure*}[t]
\centering
\includegraphics[width=0.45\textwidth]{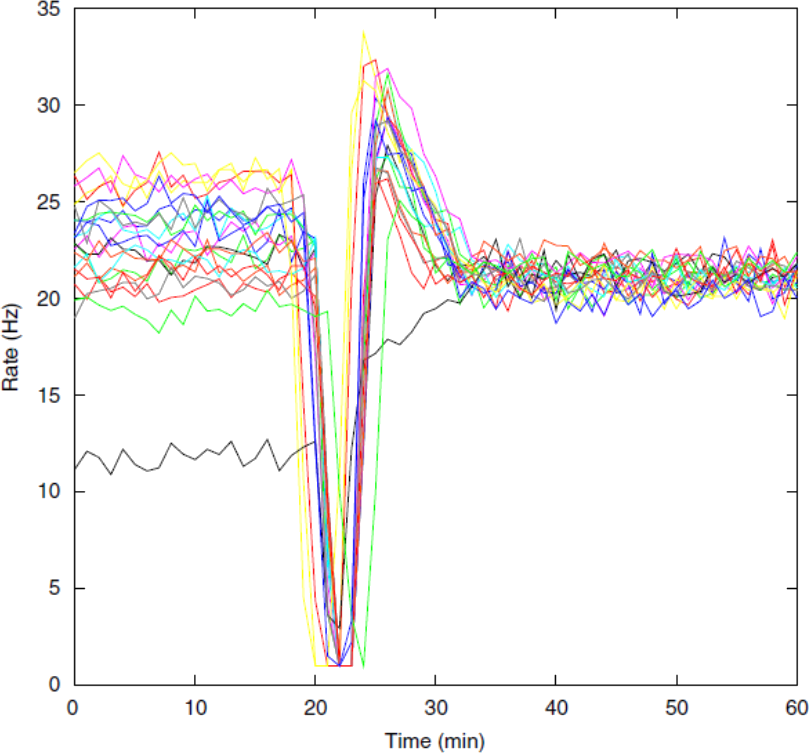}
\caption{Convergence of the 3-fold coincidence trigger at $3.2I^\text{est}_\text{VEM}$ to ${\sim}20$\,Hz after the convergence algorithm based on the $2.5I^\text{est}_\text{VEM}$ singles rate for 21 SD stations (station ID is on the right).
The convergence algorithm was turned on at $t\approx20$\,min.
The drop to 0\,Hz was caused by the re-boot of the SD stations to enable the convergence algorithm.}
\end{figure*}

\subsubsection{Pressure dependence of the on-line calibration }

The ratio of $I^\text{peak}_\text{VEM}$ to $I^\text{est}_\text{VEM}$ was found to have a very slight pressure dependence, which is expected since the online calibration uses the rate of atmospheric muons at $1.75I^\text{peak}_\text{VEM}$ to determine $I^\text{peak}_\text{VEM}$.
The dependence is clearer for technical reasons for $Q^\text{est}_\text{VEM}$/$Q^\text{peak}_\text{VEM}$, and is shown in Fig.~4.
The correlation is 0.1\% per g/cm$^2$.
The typical pressure change of the SD over one year is about 30\,g/cm$^2$, implying a maximum 3\% yearly variation in the trigger level.

\begin{figure}[t]
\centering
\includegraphics[width=\columnwidth]{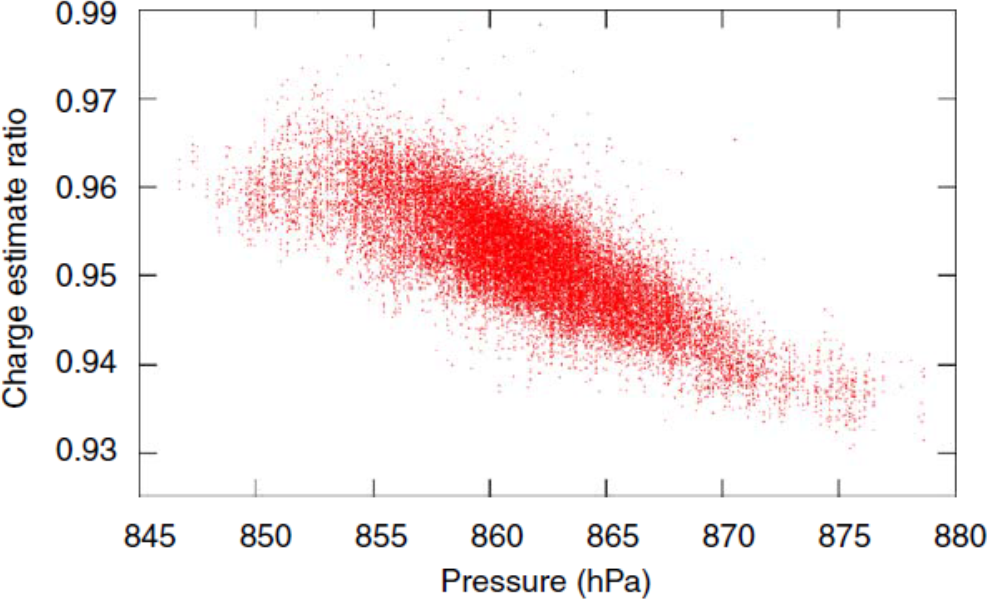}
\caption{Correlation of the ratio $Q^\text{est}_\text{VEM}$/$Q^\text{peak}_\text{VEM}$ to atmospheric pressure as measured by a weather station located at the Los Leones fluorescence site.
Note $1\,\text{hPa} = 1.020\,$g/cm$^2$ in atmospheric depth.
The effect (${\sim}0.1$\% per g/cm$^2$) on the trigger level over the course of a year is approximately 3\%.}
\end{figure}

\subsection{$Q^\text{peak}_\text{VEM}$ determination from charge histograms}

The SD electronics has a separate trigger designed specifically for collecting high-rate data with fewer bins (20 bins instead of 768 for the event data) at low threshold ($0.1I^\text{est}_\text{VEM}$).
Once the calibration procedure has stabilized $I^\text{est}_\text{VEM}$, this trigger is enabled and sets of histograms of various quantities are collected over 60\,s intervals -- approximately 150\,000 entries per histogram.
These histograms are sent to the CDAS along with any events that are requested -- therefore, each event has a high-statistics set of charge and pulse height histograms from the previous minute accompanying the data.

\begin{figure*}[t]
\centering
\includegraphics[width=0.7\textwidth]{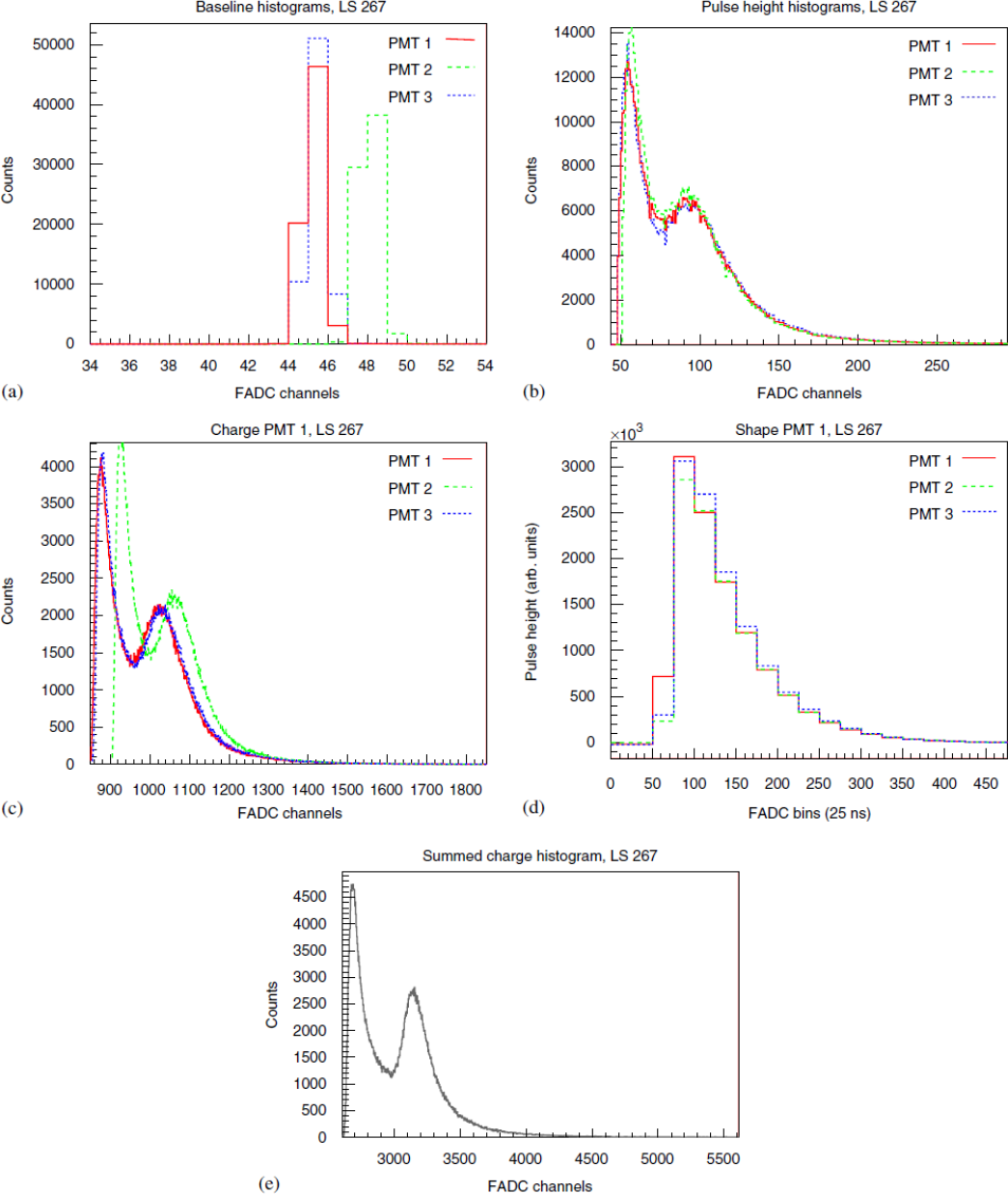}
\caption{Example calibration data which are sent back with candidate event data, built from approximately 150\,000 pulses collected in the minute prior to the event.
It includes (a) baseline histograms for all three dynode channels, (b) pulse height histograms, (c, e) charge histograms for the three PMTs and the sum of the PMTs, and (d) the pulse shape of pulses with an integrated charge of $(1.0\pm0.1)Q^\text{est}_\text{VEM}$.
The RMS spread in the $I^\text{est}_\text{VEM}$ values of the 3 PMTs was 3.3\% (the mean for a random sample of 661 tanks was 4.4\%).
The baselines are not subtracted in the charge or pulse height histograms.}
\end{figure*}

The histograms created every minute are:
\begin{itemize}
\setlength\itemsep{-1mm}
\item Charge histograms for each individual PMT. 
\item Charge histogram for the sum of all 3 PMTs.
\item Pulse height histograms for each individual PMT.
\item Histograms of the baseline of each FADC channel.
\end{itemize}

The average of all pulse shapes with an integrated charge of ($1.0\pm0.1)Q^\text{est}_\text{VEM}$ is also sent.
An example of the histograms and pulse shape average sent with each event is shown in Fig.~5.

During data analysis, the second peak of the individual charge histograms (Fig.~5c) is fit by a quadratic function to obtain the value of $Q_\text{VEM}$ used to convert the integrated signal into units of VEM.
The agreement of $Q^\text{est}_\text{VEM}$ and $Q^\text{peak}_\text{VEM}$ is a good indication that this peak is resolvable for all the SD stations.
The $Q_\text{VEM}$ obtained from charge histograms can also be cross-checked using measurements of the charge deposited in the PMT by the Cherenkov light from a decay electron from a stopped muon~[9].
These two measurements were found to agree to 0.5\%, well within the 4\% uncertainty of the $Q_\text{VEM}$ measurement from the decay electron.

\section{Additional parameters}

In addition to the primary conversion from integrated channels to VEM units, the calibration must also be able to convert the raw FADC traces into integrated channels.

There are two primary parameters needed for this.
\begin{enumerate}
\setlength\itemsep{-1mm}
\item The baselines of all six FADC inputs.
\item The gain ratio between the dynode and the anode (called the ``dynode/anode ratio'').
\end{enumerate}
The baseline is computed from each of the 100\,Hz calibration triggers obtained over a 60\,s interval (${\sim}6000$ total triggers) as well as the standard deviation of each of
the six inputs.
This information is also included with each event, and can be cross-checked against histograms of the baseline for the three dynode channels.

\subsection{Dynode/anode ratio ($D/A$)}

The dynode/anode ratio ($D/A$) is slightly more complicated to measure.
The only pulses available to measure $D/A$ are muon-like pulse shapes -- either from atmospheric muons or from an onboard LED flasher (used for linearity measurements).
A muon-like signal can be described essentially as a falling exponential after the peak, with a typical decay constant of ${\sim}60$\,ns (see Fig.~5d).
The signal/noise of the sum therefore decreases as bins farther from the peak are included in the summation.

The nominal gain between the dynode and the anode is 32 -- that is, 5 bits of overlap out of a 10\,bit FADC, giving 15 total bits of dynamic range.
For a signal which is nearly saturated on the dynode (${\sim}950$\,ch with a 50\,ch baseline), the anode signal will be merely 30\,ch above baseline.
The RMS noise of the anode and dynode channels is ${\sim}0.5-0.8$\,ch, which implies that a signal with a decay constant of ${\sim}60$\,ns will only be above the noise level on the anode for four bins (200\,ns), as compared to 17 bins for the dynode.

Ideally, the best measurement of $D/A$ would occur simply by taking the peak of the dynode divided by the peak of the anode, and averaging over many samples.
Unfortunately, the dynode signal is \emph{not} simply the anode signal multiplied by $D/A$ -- the dynode is amplified by two Analog Devices AD8012 amplifier stages, each of which has a phase delay of approximately $2-3$\,ns.
Thus the dynode is actually delayed by $4-6$\,ns, which is approximately 1/4 of a 25\,ns clock cycle.
This prevents a direct peak-to-peak comparison.
An alternative approach would be summing the signal and dividing the sum of the anode by the sum of the dynode.
This, however, is also not possible, as the error associated with the RMS noise of the dynode and anode becomes quite significant.
Structured noise (below the RMS noise level) due to channel-to-channel crosstalk or other temporally correlated noise sources prevents obtaining an accurate (${<}$5\%) $D/A$ measurement even with large statistics.

$D/A$ is therefore measured by modelling the anode signal shape ($A$) from the dynode ($D$) as
\begin{equation}
A(t) = \frac{1}{R}((1-\varepsilon)D(t) + \varepsilon D(t+1))
\end{equation}
where $t$ is the time bin, $R$ is $D/A$, and $\varepsilon$ is the fractional bin offset of the dynode.
$R$ and $\varepsilon$ are determined using $\chi^2$ minimization.
$D$ and $A$ are determined with about 100 pulses taken within 3\,min.
Here, $\varepsilon$ is known to be positive, but is allowed to vary for the fit.
This procedure also has the advantage of measuring the phase delay and any time dependence it may have.

An example of the fit (and the fit region) is shown in Fig.~6, for a version of the front end electronics with higher noise characteristics than the production version.
The two pulses were generated by a resistively-divided pulse generator, which had no phase delay between the dynode and the anode.
For these pulses, the ``dynode/anode'' ratio was 33.9 by design, and measured independently with an oscilloscope.
This procedure gave a dynode/anode ratio of 33.3, accurate to within 2\%.
The fitting procedure correctly gave a very small value for $\varepsilon$ for this fit (${<}10^{-6})$.
For actual stations, however, $\varepsilon$ is measured to be on average 0.23, with an individual precision of 0.04.
This corresponds to a delay of 5.8\,ns, in agreement with our expectations.
$D/A$ determined with this method are also in good agreement with direct measurements of the $D/A$ on the PMT base.
However, estimating the accuracy of this comparison is quite difficult as the dependence of $D/A$ on the high voltage of the PMT is poorly measured.

\begin{figure*}[t]
\centering
\includegraphics[width=0.5\textwidth]{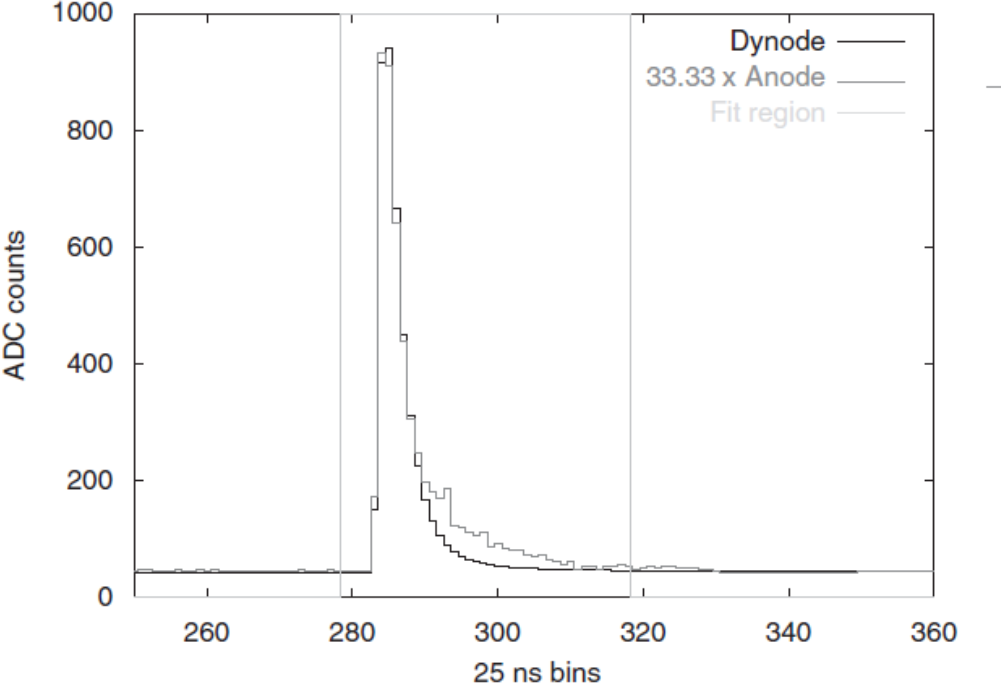}
\caption{Example of the $D/A$ fit to a resistively-divided pulse with the preproduction front end electronics.
The pulse generator had an intrinsic $D/A$ of 33.9, and the $D/A$ fit method gave a $D/A$ of 33.3, within 2\%.
The excess in the anode seen near 300 time bins is ${\sim}1$\,ch, which is the noise level of the preproduction electronics.
The fit clearly determines the proper $D/A$ even in the presence of ${<}1$\,ch noise.}
\end{figure*}

\section{Conclusion}

The main calibration goal for the surface detector is to convert the integrated flash ADC signal into vertical equivalent muon (VEM) units, and to provide a stable and uniform trigger for the detector.
The conversion to VEM units is done by determining $Q_\text{VEM}$ through their relation to a peak in charge ($Q_\text{VEM}^\text{peak}$) histograms, which is determined through an independent experiment.
$Q_\text{VEM}^\text{peak}$ is measured with a high-statistics (150\,000 entries) charge histogram every minute, and agrees with an independent local software estimate to 3\%.

Conversion of the anode signal requires the determination of the dynode/anode ratio ($D/A$), which is done by averaging large pulses and performing a linear time-shifted fit to determine both the $D/A$ and the phase delay between the two signals.
This method, when performed on two resistively divided signals determined the $D/A$ to 2\%.

Uniform trigger levels are provided by estimating $I_\text{VEM}^\text{peak}$ -- the peak in a pulse height histogram -- via a $\sigma$-$\delta$ convergence algorithm on a 70\,Hz singles rate inside a 100\,Hz 3-fold coincidence.
This measurement is precise to 6\%.
The use of a rate to determine $I_\text{VEM}^\text{peak}$ introduces a small systematic pressure dependence in the trigger level of approximately 0.1\% per g/cm$^2$ leading to less than a 3\% effect over the course of a year.

The calibration parameters mentioned here are determined every 60\,s and returned to the central data acquisition system (CDAS) with each event and stored along with the event data.
Each event therefore contains a large amount of information about the state of each surface detector in the minute preceding the trigger, allowing for an accurate calibration of the data.

\section*{References}
 
\begin{sloppypar}
\footnotesize\setlength{\parindent}{0mm}\setlength{\parskip}{2mm}

\noindent
[1] M.~Lawrence, R.J.O.~Reid, and A.A.~Watson, J.~Phys.~G 17 (1991) 733.

\noindent
[2] J.~Abraham et al., Nucl.~Instr.~Meth.~A 523 (2004) 50.

\noindent
[3] P.D.J.~Clark and D.~Nitz, Proceedings of 27th ICRC, Hamburg, vol.~2,
2001, p.~765.

\noindent
[4] D.~Allard et al., Proceedings of 29th ICRC, Pune, vol.~7, 2005, p.~71.

\noindent
[5] P.~Bauleo et al., Nucl.~Instr.~Meth.~A 406 (1998) 69.

\noindent
[6] M.~Aglietta et al., Proceedings of 29th ICRC, Pune, vol.~7, 2005, p.~83.

\noindent
[7] A.~Etchegoyen et al., Nucl.~Instr.~Meth.~A 545 (2005) 602.

\noindent
[8] D.~Nitz for the P.~Auger Collaboration, IEEE Trans.~Nucl.~Sci.~NS-51
(2004) 413.

\noindent
[9] P.~Allison et al., Proceedings of 29th ICRC, Pune, vol.~8, 2005, p.~299.
\end{sloppypar}

\end{document}